\def\d{\partial}
\def\dh{\mathop{\vphantom{\odot}\hbox{$\partial$}}}
\def\dl{\dh^\leftrightarrow}
\def\sqr#1#2{{\vcenter{\vbox{\hrule height.#2pt\hbox{\vrule width.#2pt 
height#1pt \kern#1pt \vrule width.#2pt}\hrule height.#2pt}}}}
\def\w{\mathchoice\sqr45\sqr45\sqr{2.1}3\sqr{1.5}3\,}

\def\psq{{\overline{\psi}}}

\def\=d{\,{\buildrel\rm def\over =}\,}

\def\i3p{\p32\int d^3p}

\def\As{A\hbox to 1pt{\hss /}}
\def\np4{\int d^4p_1\cdots d^4p_{n-1}\, }

\def\nx4{\int d^4x_1\ldots d^4x_n\, }

\def\kon#1#2{\vbox{\halign{##&&##\cr
\lower4pt\hbox{$\scriptscriptstyle\vert$}\hrulefill &
\hrulefill\lower4pt\hbox{$\scriptscriptstyle\vert$}\cr $#1$&
$#2$\cr}}}

\def\konv#1#2#3{\hbox{\vrule height12pt depth-1pt}
\vbox{\hrule height12pt width#1cm depth-11.6pt}
\hbox{\vrule height6.5pt depth-0.5pt}
\vbox{\hrule height11pt width#2cm depth-10.6pt\kern5pt
      \hrule height6.5pt width#2cm depth-6.1pt}
\hbox{\vrule height12pt depth-1pt}
\vbox{\hrule height6.5pt width#3cm depth-6.1pt}
\hbox{\vrule height6.5pt depth-0.5pt}}
\def\konu#1#2#3{\hbox{\vrule height12pt depth-1pt}
\vbox{\hrule height1pt width#1cm depth-0.6pt}
\hbox{\vrule height12pt depth-6.5pt}
\vbox{\hrule height6pt width#2cm depth-5.6pt\kern5pt
      \hrule height1pt width#2cm depth-0.6pt}
\hbox{\vrule height12pt depth-6.5pt}
\vbox{\hrule height1pt width#3cm depth-0.6pt}
\hbox{\vrule height12pt depth-1pt}}

\def\konw#1#2#3{\hbox{\vrule height12pt depth-1pt}
\vbox{\hrule height12pt width#1cm depth-11.6pt}
\hbox{\vrule height6.5pt depth-0.5pt}
\vbox{\hrule height12pt width#2cm depth-11.6pt \kern5pt
      \hrule height6.5pt width#2cm depth-6.1pt}
\hbox{\vrule height6.5pt depth-0.5pt}
\vbox{\hrule height12pt width#3cm depth-11.6pt}
\hbox{\vrule height12pt depth-1pt}}

\def\i{{\rm int}}

\def\m3{{\mu_1\mu_2\mu_3}}

\def\p{{(+)}}

\documentclass{article}
\usepackage{amssymb}
\begin{document}
\title{Deformation stability of BRST-quantization}
\author{M. D\"utsch\thanks{Work supported by the Alexander von Humboldt 
Foundation}  and K. Fredenhagen\\[1mm]
Institut f\"ur Theoretische Physik\\
Universit\"at Hamburg\\
149, Luruper Chaussee\\
D-22761 Hamburg, Germany\\
{\tt duetsch@mail.desy.de, fredenha@@x4u2.desy.de}}

\date{}

\maketitle

\begin{abstract}
To avoid the problems which are connected with the long distance 
behavior of perturbative gauge theories we present a {\it local} 
construction of the 
observables which does not involve the adiabatic limit. First 
we construct
the interacting fields as formal power series by means of causal 
perturbation theory. The observables are defined by BRST invariance
where the 
BRST-transformation $\tilde s$ acts as a graded derivation on the algebra 
of interacting
fields. Positivity, i.e. the existence of Hilbert space representations
of the local algebras of observables is shown with the help of 
a local Kugo-Ojima operator
$Q_{\rm int}$ which implements $\tilde s$ on a local algebra 
and
differs from the corresponding operator $Q$ of the free theory. We prove
that the Hilbert space structure present in the free case is stable under
perturbations. All assumptions are shown to be satisfied in QED in a
finite spatial volume with suitable boundary conditions. As a 
by-product
we find that the BRST-quantization is not compatible with periodic boundary
conditions for massless free gauge fields.\\

perturbation theory 

\end{abstract}

\section*{1. Introduction}
The long distance behavior of nonabelian perturbative gauge theories 
is plagued by serious problems. In {\it massless} theories there 
appear infrared divergences in the adiabatic limit
$g\rightarrow$ const. of the S-matrix, where $g$ is a space-time dependent
coupling 'constant'. In QED these divergences are logarithmic and cancel in 
the cross section. (This is proven at least at low orders of the perturbation 
series \cite{S}.) Moreover, Blanchard and Seneor \cite{BlSe} 
proved that the adiabatic 
limit of Green's and Wightman functions exists for QED.
But in nonabelian gauge theories the divergences are worse. Perturbation 
theory seems to be not able to describe the long distance properties of these 
models ("confinement"). In {\it massive} theories the infrared divergences 
are absent, but e.g. in the electroweak theory an S-matrix formalism suffers
from the instability of some particles, e.g. the W-, Z-bosons and the muons 
and taus. States containing such particles belong to the physical state 
space, but they cannot appear as asymptotic states of the S-matrix for 
$t\rightarrow\pm\infty$.

Our way out is to {\it construct the observables locally}. 
We consider a fixed, open double cone ${\cal O}
\subset {\bf R}^4$. The coupling 'constant' $g$ has compact support and
takes a constant value on ${\cal O}$
$$g\in {\cal D}({\bf R}^4),\quad\quad\quad g(x)=e={\rm const.},\quad\forall
x\in {\cal O}.\eqno(1.1)$$

The {\it interacting fields} are defined by Bogoliubov's formula \cite{BS}
$$A_{{\rm int}\>{\cal L}}(x)\=d {\delta\over i\delta h(x)}S({\cal L})^{-1}
S({\cal L}+hA)\vert_{h=0},\eqno(1.2)$$
and the time ordered products $T({\cal L}(x_1)...{\cal L}(x_n))$,
which appear in the S-matrix
$$S({\cal L})=\sum_{n=0}^\infty {i^n\over n!}\int d^4x_1...d^4x_n\,
T({\cal L}(x_1)...{\cal L}(x_n)),\eqno(1.3)$$
are constructed by means of  causal perturbation theory \cite{BS,BF,EG,S,St2}.
The interacting fields $A_{{\rm int}\>{\cal L}}(x)$ ($A$ is a Wick polynomial
of incoming fields) are formal power series of operator 
valued distributions on a
dense invariant domain ${\cal D}$ in the Fock space of incoming fields.
They depend on an interaction Lagrangian ${\cal L}$
which is a Wick polynomial of incoming fields with testfunctions $g\in {\cal D}
({\bf R}^4)$ as coefficients.

The crucial observation is that the dependence of the interacting fields on the
interaction Lagrangian is {\it local}, in the sense that
Lagrangians ${\cal L}_1$ and
${\cal L}_2$ which differ only within a closed region which does not
intersect the closure of ${\cal O}$, lead to unitarily equivalent fields
within ${\cal O}$,
i.e. there exists a unitary formal power series $V$ of operators on ${\cal D}$
such that
$$VA_{{\rm int}\>{\cal L}_1}(x)V^{-1}=A_{{\rm int}\>{\cal L}_2}(x),
\quad\quad\quad
\forall x\in {\cal O},\eqno(1.4)$$ 
and $V$ does not depend on $A$ \cite{BF}. The proof of (1.4) 
relies on the causal 
factorization of the time ordered products. 

The field algebra $\tilde {\cal F}({\cal O})$ which
is generated by
$$\{A_{{\rm int}\>{\cal L}}(f)=\int d^4x\,
A_{{\rm int}\>{\cal L}}(x)f(x)\vert f\in {\cal D(O)}\}\eqno(1.5)$$  
is up to unitary equivalence uniquely determined by
$g\vert_{\cal O}$. Since ${\cal O}$ is arbitrary, the full net of local 
algebras can be constructed without ever performing the adiabatic limit 
$g\rightarrow {\rm constant}$.

In gauge theories the (local)
algebras of interacting fields contain unphysical fields like vector
potentials and ghosts.
They can be eliminated by the BRST formalism. But it remains to show 
that the algebra of observables can be (nontrivially) 
represented on a Hilbert space.

In the free theory positivity can be  verified by an explicit 
calculation. Formally, in the adiabatic limit, the positivity is hence valid
also for the interacting theory. 
We show, that for a localized interaction (i.e. before the adiabatic
limit), the physical Hilbert space
can be obtained as a deformation of the free one
(sect. 2). The construction relies on some assumptions, which are verified for
the example of QED (sect. 3). To avoid a volume divergence in $Q_{\rm int}$
we embed our double cone ${\cal O}$ into the cylinder ${\bf R}\times C_L$,
where ${\bf R}$ denotes the time axis and $C_L$ is a cube of length $L$.
In sect. 4 we point out the importance of a suitable choice of boundary 
conditions for the BRST-quantization of massless free gauge
fields on ${\bf R}\times C_L$.

We hope that due to its local character, our construction can be 
generalized to curved space-times, continuing the program of \cite{BF,BFK}.

\section*{2. Connection of observable algebras and field algebras in
perturbative gauge theories}

\subsection*{2.1 Local construction of observables in gauge theories 
and representation in the physical pre Hilbert space}

Let ${\cal F}$ be a ${\bf Z}_2$-graded *-algebra, e.g.  
the algebra of fields of a gauge theory where the ${\bf Z}_2$-gradiation is 
$(-1)^{\delta (F)}$ with the ghost
number $\delta (F)$.
To get rid of the unphysical fields,  
we use the BRST-transformation 
$s$ \cite{BRS}, which is a graded 
derivation on 
${\cal F}$ with $s^2=0$
and $s(F^*)=-(-1)^{\delta (F)}s(F)^*$.
 
The kernel of
$s$, ${\cal A}_0:=s^{-1}(0)$, 
is a $*$-subalgebra of ${\cal F}$ and  
${\cal A}_{00}:=s({\cal F})$ is a 2-sided ideal in ${\cal A}_0$.
Hence we may define the {\it algebra of observables} as the quotient
$${\cal A}\=d {{\cal A}_0\over {\cal A}_{00}}. \eqno(2.1)$$ 

We ask now under which conditions ${\cal A}$ has a 
nontrivial *-representation
by operators on a pre Hilbert space.
For this purpose we work with the Kugo-Ojima formalism \cite{KO}. We
assume that ${\cal F}$ has a faithful representation on 
an inner product space $({\cal K},<.,.>)$ such that
$<F^*\phi,\psi>=<\phi,F\psi>,\>\forall F\in {\cal F}$,
and that $s$ is implemented by an operator $Q$ on ${\cal K}$, i.e.
$$s(F)=QF-(-1)^{\delta (F)}FQ,\eqno(2.2)$$
such that
$$<Q\phi,\psi>=<\phi,Q\psi>\quad\quad\quad {\rm and}\quad\quad\quad Q^2=0.
\eqno(2.3)$$
Note that if the inner product on ${\cal K}$ is positive definite, we find
$<Q\phi,Q\phi>=\break <\phi,Q^2\phi>=0$, hence $Q=0$ and thus also $s=0$. Hence
for nontrivial $s$ the inner product must necessarily be indefinite.
 
Let ${\cal K}_0\=d {\rm Ke}\,Q$ be the kernel 
and  ${\cal K}_{00}$ 
the range of $Q$.
Because of $Q^2=0$ we have ${\cal K}_{00}\subset {\cal K}_0$. We assume:
$${\bf (Positivity)}\quad\quad\quad {\rm (i)}\quad <\phi,\phi>\geq 0
\quad\quad\forall\phi\in {\cal K}_0,$$
$$\quad\quad\quad {\rm and}\quad\quad {\rm (ii)}\quad\phi\in {\cal K}_0\quad 
\wedge\quad <\phi,\phi>=0\quad\Longrightarrow
\quad \phi\in {\cal K}_{00}.\eqno(2.4)$$

Then 
$${\cal H}\=d {{\cal K}_0\over {\cal K}_{00}},\quad\quad\quad
<[\phi_1],[\phi_2]>_{\cal H}:\=d <\psi_1,\psi_2>_{\cal K},
\quad\psi_j\in [\phi_j]:=\phi_j+{\cal K}_{00}\eqno(2.5)$$
is a pre Hilbert space and 
$$\pi ([A])[\phi]\=d [A\phi]\eqno(2.6)$$ 
is a representation on ${\cal H}$ (where $A\in {\cal A}_0,\,\phi\in 
{\cal K}_0,\,[A]:=A+{\cal A}_{00}$) \cite{DF}.

\subsection*{2.2 Stability under deformations}

It is gratifying that the described structure is {\it stable under 
deformations}, e.g. by turning on the interaction.
Let ${\cal K}$ be fixed and replace $F\in {\cal F}$ by a formal power series
$\tilde F=\sum_n g^nF_n$ with $F_0=F$ and $F_n\in {\cal F},\,
\delta(F_{n})={\rm const}$. In the same
way replace $s$ and $Q$ by formal power series $\tilde s=\sum_n g^ns_n$,
$\tilde Q=\sum_n g^nQ_n$ with $s_0=s$, $Q_0=Q$ and
$$\tilde s^2=0,\quad\tilde Q^2=0,\quad <\tilde Q\phi,\psi>=<\phi,
\tilde Q\psi>\quad {\rm and}\quad \tilde s(\tilde F)=\tilde Q\tilde F
-(-1)^{\delta (\tilde F)}\tilde F\tilde Q.\eqno(2.7)$$
We can then define $\tilde {\cal A}\=d {{\rm Ke}\,\tilde s\over 
{\rm Ra}\,\tilde s}$. ${\cal K}_0$ and ${\cal K}_{00}$
have to be replaced by formal power series $\tilde {\cal K}_0:=
{\rm Ke}\,\tilde Q$ and $\tilde {\cal K}_{00}:={\rm Ra}\,\tilde Q$
with coefficients in ${\cal K}$. Due to the above result (2.6), the algebra
$\tilde {\cal A}$ has a natural representation on 
$\tilde {\cal H}\=d {\tilde {\cal K}_0\over \tilde {\cal K}_{00}}$.
The inner product on ${\cal K}$ induces an inner product 
on $\tilde{\cal H}$ which assumes values in the formal 
power series over ${\bf C}$. 
We adopt the point of view that a formal power series $\tilde b=\sum_ng^nb_n,
\>b_n\in {\bf C}$ is {\it positive} if there is another formal power series
$\tilde c=\sum_ng^nc_n,\>c_n\in {\bf C}$ with $\tilde c^*\tilde c=\tilde b$,
i.e. $b_n=\sum_{k=0}^n \bar c_k c_{n-k}$. (cf. \cite{BW})

The assumptions concerning the positivity of the inner
product are automatically fulfilled for the deformed theory, 
if they hold true in the undeformed model \cite{DF}.

{\bf Theorem 1:} Let the positivity assumption (2.4) be fulfilled in zeroth 
order. Then

(i) $<\tilde\phi,\tilde\phi>\geq 0\quad\quad\forall 
\tilde\phi\in \tilde {\cal K}_0$,

(ii) $\tilde\phi\in \tilde {\cal K}_0\quad \wedge\quad <\tilde\phi,
\tilde\phi>=0\quad\Longrightarrow\quad \tilde\phi\in \tilde {\cal K}_{00}.$

(iii) For every $\phi\in {\cal K}_0$
there exists a power series $\tilde\phi\in\tilde {\cal K}_0$ with 
$(\tilde\phi)_0=\phi$.

(iv) Let $\pi$ and $\tilde\pi$ be the representations (2.6) of 
${\cal A},\,\tilde {\cal A}$ on ${\cal H},\,\tilde {\cal H}$ 
respectively. Then
$\tilde{\pi}(\tilde { A})\neq 0$ if $\pi(A_{0})\neq 0$.

From parts (i) and (ii) we conclude that 
$\tilde {\cal H}= {\tilde {\cal K}_0\over \tilde {\cal K}_{00}}$ 
is a perturbative
analog of a (pre) Hilbert where the scalar product assumes values in 
the formal power 
series over ${\bf C}$.
Note that $\phi\rightarrow\tilde\phi$ is non-unique and this
holds also true for the  induced relation between ${\cal 
H}$ and $\tilde {\cal H}$.
A consequence of part (i) of the theorem is the {\it positivity of the 
Wightman distributions of $\tilde s$-invariant fields} \cite{DF}.

\section*{3. Verification of the assumptions in models}
 
Kugo-Ojima \cite{KO} argue that at 
asymptotically early times 
the interacting fields tend to the free incoming fields. Since 
$Q_{\rm int}\=d \tilde Q$ is conserved, it coincides with 
the free one $Q=Q_0$. Hence it is sufficient to 
check the assumptions for the free theory. But the BRST-current (i.e. the 
current belonging to $Q_{\rm int}$) is only conserved in regions 
where $g$ is constant (see below and \cite{BDF}). Hence,
the Kugo-Ojima procedure involves the (partial) adiabatic 
limit for $t\rightarrow
-\infty$, which is difficult to control. The argument does certainly 
not work in nonabelian gauge theories (as can be seen by
an explicit calculation of the first order of $Q_{\rm int}$) \cite{BDF}.
We therefore prefer not to work in the adiabatic limit. The 
price to pay is that $Q_{\rm int}$ does  not agree with $Q$, 
so for the construction of the physical Hilbert space we 
have to check the assumptions of section 2. We do this for QED.
We see no principle obstacle for the generalization to nonabelian gauge 
theories. But the details still need to be worked out \cite{BDF}.

\subsection*{3.1 Free QED}

The field algebra ${\cal F}$ is generated by the free photon fields $A^\mu$
in Feynman gauge, the free spinor fields
$\psi$ and $\psq$, a pair of free ghost fields
$u$ and $\tilde u$, the Wick monomials 
$j^\mu =:\psq\gamma^\mu\psi:\, ,\,\gamma_\mu A^\mu\psi,\,\psq\gamma_\mu 
A^\mu,\,j_\mu A^\mu$ and the derivated free fields $\d_\mu A^\mu$,
$F^{\mu\nu}=\d^\mu A^\nu-\d^\nu A^\mu$.  
This algebra is faithfully 
represented on a Krein space which is given by the usual Fock space of 
free fields and a Krein operator which defines the indefinite inner product.
The graded derivation $s$ is determined by the 
BRST-transformation of free fields
$$s(A^\mu)=i\d^\mu u,\quad s(\psi)=0,\quad s(\psq)=0,\quad s(u)=0,
\quad s(\tilde u)=-i\d_\mu A^\mu\eqno(3.1)$$
and by translation invariance of $s$.
This transformation is implemented by the free Kugo-Ojima charge \cite{DHKS}
$$Q\=d \int_{x_0={\rm const.}} d^3x\,(\d_{\nu}A^{\nu}(x)){\dl}_0u(x),
\eqno(3.2)$$
which fulfills\footnote{We restrict all operators 
(resp. formal power series of operators)
to the dense invariant domain ${\cal D}$ and, therefore, there is no difference
between symmetric and self-adjoint operators.} $Q^*=Q$,
and $Q^2=0$. In addition the inner product
$<.,.>$ is positive semidefinite on ${\rm Ke}\,Q$ and the 
space of nullvectors in ${\rm Ke}\,Q$ is precisely ${\rm Ra}\,Q$ (\cite{DHS,K})

\subsection*{3.2 Construction of the interacting Kugo-Ojima charge in QED}

In QED the interaction is given by
$${\cal L}(x)=g(x):\psq (x)\gamma_\mu A^\mu (x)\psi (x):,\quad
\quad g\in {\cal D}({\bf R}^4).\eqno(3.3)$$
We fix the double cone ${\cal O}$ to be the causal completion of the surface
$\{ (0,{\bf x}),|{\bf x}|< r\}$ and assume 
the switching function $g\in {\cal D}({\bf R}^4)$ to be constant on 
a neighbourhood ${\cal U}$ of $\bar {\cal O}$
(1.1). We study the algebra $\tilde {\cal F}({\cal O})$ (1.5) of interacting
fields localized in ${\cal O}$. The ghost fields do not couple in QED, hence 
$u_{{\rm int}\>{\cal L}}(x)=u(x)$ and
$\tilde u_{{\rm int}\>{\cal L}}(x)=\tilde u(x)$
The interacting fields can be normalized such that they fulfil the {\it field 
equations} \cite{DF,DKS}
$$\w A_{{\rm int}\>{\cal L}}^\mu (x)=-g(x)j_{{\rm int}\>{\cal L}}^\mu (x),
\eqno(3.4)$$
$$(i\gamma_\mu \d^\mu -m)\psi_{{\rm int}\>{\cal L}}(x)=
-g(x)(\gamma_\mu A^\mu \psi)_{{\rm int}\>{\cal L}}(x),\eqno(3.5)$$
electric current conservation
$$\d_\mu j^\mu_{{\rm int}\>{\cal L}}=0\eqno(3.6)$$
and the following commutation relations 
at points $x,y\in {\cal O}$ \cite{DF} 
$$[\d_\mu A^\mu_{{\rm int}\>{\cal L}}(x),A^\nu_{{\rm int}\>{\cal L}}(y)]=
i\d^\nu D (x-y),\quad\quad
[\d_\mu A^\mu_{{\rm int}\>{\cal L}}(x),\d_\nu A^\nu_{{\rm int}
\>{\cal L}}(y)]=0,
\eqno(3.7)$$
$$[\d_\mu A^\mu_{{\rm int}\>{\cal L}}(x),\psi_{{\rm int}\>{\cal L}}(y)]=
D (x-y)e\psi_{{\rm int}\>{\cal L}}(y),\eqno(3.8)$$
where $D$ is the massless Pauli-Jordan distribution.

The abelian BRST-transformation $\tilde s=s_0+gs_1$ \cite{BRS} is
a graded $*$-derivation with zero square
which induces the following transformations on the basic fields,
$$\tilde s(A^\mu_{{\rm int}\>{\cal L}}(x))=i\d^\mu u(x),
\quad\quad \tilde s(u(x))=0,
\quad\quad \tilde s(\tilde u(x))=-i\d_\mu A^\mu_{{\rm int}\>{\cal L}}(x),$$
$$\tilde s(\psi_{{\rm int}\>{\cal L}}(x))=
-e\psi_{{\rm int}\>{\cal L}}(x)u(x),\quad\quad\tilde
s(\psq_{{\rm int}\>{\cal L}}(x))=e\psq_{{\rm int}\>{\cal L}}(x)u(x)
\eqno(3.9)$$
for $x\in {\cal O}$. (The pointwise products above are well defined.)
 
On $\tilde {\cal F}({\cal O})$ $\tilde s$ is implemented 
by the operator
$$Q_{\rm int}(g,k)=\int d^4x\,k(x)(\d_{\nu}A^{\nu}_{{\rm 
int}\>{\cal L}}(x)){\dl}_0^x u(x)\eqno(3.10)$$ 
(where $k\in{\cal D}({\cal U})$ is 
a suitably chosen smeared characteristic function
of the surface $\{ (0,{\bf x}),|{\bf x}|\le r\}$).
Note that $[Q_{\rm int}(g),F]_\mp,\> F\in
\tilde {\cal F}({\cal O})$ is independent of
$k$, since the BRST-current 
$\d_\mu A^\mu_{{\rm int}\>{\cal L}}(x) {\dl}_\nu^x u(x)$ is conserved
within $\cal U$. 
$Q_{\rm int}(g,k)$ is hermitian for real valued $k$
and nilpotent,
$$Q_{\rm int}(g,k)^2={1\over 2}\{Q_{\rm int}(g,k),Q_{\rm int}(g,k)\}=$$
$$={1\over 2}\int d^4x\,h(x)\int d^4y\,h(y)
[\d_\mu A^\mu_{{\rm int}\>{\cal L}}(x),\d_\nu A^\nu_{{\rm int}\>{\cal L}}(y)]
{\dl}_0^x {\dl}_0^y u(x)u(y)=0,\eqno(3.11)$$
by means of (3.7).

But we need in addition that the zeroth order term $Q_0(k)$ of $Q_{\rm 
int}(g,k)$ (3.10) satisfies the positivity assumption 
(2.4). There seems to be no reason why this should 
hold for a generic choice of $k$. One might try to control 
the limit when $k$ tends to a smeared characteristic 
function of the $t=0$ hyperplane (in order that $Q_0(k)$ becomes equal to
the free charge Q (3.2)), but without an a priori information on the existence
of an $\tilde s$-invariant state this appears to be a hard problem.    
 
There is a more elegant way to get rid of these problems 
which relies on the local character of our construction. We 
may embed our double cone  ${\cal O}$ isometrically into 
the cylinder  ${\bf R}\times C_L$, where $C_L$ is a cube of 
length $L,\quad L\gg r$, with suitable boundary 
conditions (see sect. 4), and where the first factor denotes the time 
axis. If we choose the compactification length $L$ 
big enough, the
properties of the local algebra $\tilde {\cal F}({\cal O})$ are not changed.

We assume the switching function $g$ to fulfil
$$g(x)=e={\rm constant}\quad\quad\forall x\in {\cal O}\cup
\{(x_0,\vec x)|\,|x_0|<\epsilon\}\quad\quad (r\gg\epsilon>0)\eqno(3.12)$$
on ${\bf R}\times C_L$ and to have compact support in 
timelike directions. Now we may insert 
$$k(x):= h(x_0),\quad\quad {\rm where}\quad\quad 
h\in {\cal D}([-\epsilon,\epsilon]),
\quad\quad\int dx_0\,h(x_0)=1\eqno(3.13)$$
into the expression
(3.10) for $Q_{\rm int}$. 
The zeroth order
$Q_0$ then agrees with the free charge $Q$ on $C_L$ (3.2), hence we
may apply Theorem 1. 

We emphasize that our construction shall describe QED also in the {\it
non-compactified} Minkowski space (this is the main concern of the paper)
and, therefore, should not depend on the compactification
length $L$. On the level of the algebras this is evident. 
We conjecture that also the state space
(i.e. the set of expectation functionals induced by vectors in the
physical Hilbert space) is independent of $L$, but this remains to be proven.

An open question is the {\it physical meaning of the remaining 
normalization conditions} in a local perturbative 
construction, after the restrictions from gauge invariance and other
symmetries were taken into account.
The parameters involved may be considered as structure constants of the 
algebra of observables, but their usual interpretation as charge and mass
involve the adiabatic limit.

\section*{4. Boundary conditions for massless free gauge fields
in a finite volume}

The purpose of this section is to demonstrate the {\it importance of a 
suitable choice of boundary conditions}. First we show that the 
BRST-quantization 
is not compatible with periodic boundary conditions for massless
free gauge fields. 

Let $T_3$ be the 3-torus of
length $L$. The algebra of a free massless scalar 
field $\varphi$  on ${\bf R}\times T_3$ with 
periodic boundary 
conditions is the unital *-algebra generated by elements 
$\varphi(f)$, $f\in {\cal D}({\bf R}\times T_3)$ with the 
relations 
$$f\mapsto\varphi(f) {\rm \ is \ linear },\eqno(4.1)$$
$$\varphi(\w f)=0,\eqno(4.2)$$
$$\varphi(f)^{*}=\varphi(\bar f),\eqno(4.3)$$
$$[\varphi(f),\varphi(g)]=\int 
d^4xd^4yf(x)g(y)D_{L}(x,y),\eqno(4.4)$$
where $D_{L}$ is the fundamental solution of the wave 
equation on ${\bf R}\times T_3$ with periodic boundary 
conditions, which has the explicit form
$$D_{L}(x^0,{\vec x},y^0,{\vec y})=\sum_{{\vec{n}}\in 
{\bf Z}^3}D(x^0-y^0,{\vec x}-{\vec y}-{\vec n}).\eqno(4.5)$$
In particular one sees that $D_{L}$ coincides with 
$D$ (the massless commutator function on Minkowski space) 
on ${\cal O}$ if the closure of the double cone 
${\cal O}$ is contained in ${\bf R}\times T_3$, considered 
as a region in Minkowski space. Hence the algebra ${\cal F}
({\cal O})$ associated 
to ${\cal O}$ is independent of the boundary conditions.

In a mode decomposition of $D_{L}$,
$$D_L(x)={i\over 2L^3}\sum_{\vec n\in {\bf Z}^3,\vec n\not= \vec 0}{1\over
\omega_{\vec n}}(e^{-i\omega_{\vec n}x^0}-e^{i\omega_{\vec n}x^0})
e^{i\vec k_{\vec n}\vec x} +{x_0\over L^3},\eqno(4.6)$$
the zero mode plays a special role.\footnote{To verify (4.6) note
that it is a solution of the wave equation and has the same 
Cauchy data as (4.5) for $x_0=0$.} 
The zero mode part of $\varphi$ (4.1-4) is defined by
$$\varphi_0(t)\=d {1\over L^3}
\int_{T_3}d^3x\,\varphi (t,\vec x).\eqno(4.7)$$
The algebra of the zero mode is isomorphic to the algebra of $p$ and 
$q$ in quantum mechanics with the free time evolution 
$\varphi_0(t)=q+pt$. There exists no ground state on this algebra.  

In Feynman gauge, the components of the photon field 
$A_{\mu}$ are quantized as scalar fields, with a minus sign 
for the commutator of the zero component. The zero mode of 
the field $\d_\mu A^\mu$ is then $-p^0L^{-3}$, which has a trivial 
kernel. This makes it 
impossible to impose the Gupta-Bleuler condition on the 
physical state space. 

The BRST formalism is even worse. The ghost fields are quantized by
$$(u(f)+i\tilde u(g))^2=\int d^4x\,d^4y\,f(x)g(y)
D_L(x-y).\eqno(4.8)$$
Inserting $f(x)=L^{-3}\delta(x_0-t_1)$ and $g(y)=-L^{-3}
\delta^\prime (y_0-t_2)$ we obtain
$$(u_{0}(t_1)+i\d_{0}\tilde u_{0}(t_2))^2=-L^{-3}\eqno(4.9)$$
for the zero mode parts. $u_{0}$ and $\d_{0}\tilde u_{0}$ are
BRST invariant, hence they are observables. In addition
$(u_{0}(t_1)+i\d_{0}\tilde u_{0}(t_2))$ is hermitian. We conclude
that there is no nonzero (pre) Hilbert space representation 
of the algebra of observables.

$(u_{0},\d_{0}\tilde u_{0})$ corresponds to a 'singlet pair' in the
terminology of \cite{KO}, sect. 3.1. Already there it was pointed out that 
the appearance of such a pair makes a consistent formulation impossible.

The way out is to choose {\it boundary conditions which exclude the zero 
mode}. For the electromagnetic field we may use metallic boundary 
conditions, i.e. the pullback of the 2-form $F$ vanishes at 
the boundary (which means that the tangential components of 
the electric field 
and the normal component of the magnetic field vanish). In 
addition we assume that the auxiliary Nakanishi-Lautrup 
field $B=\d^{\mu}A_{L\,\mu}$ (in Feynman gauge) satisfies 
Dirchlet boundary conditions. Also
the ghost and antighost fields are quantized with Dirichlet boundary
conditions. The details are worked out in appendix A of \cite{DF}.

The BRST-quantization requires no restrictions on the boundary 
conditions for the electron field. For simplicity, we choose 
periodic boundary conditions. They have the big advantage that
they are invariant under charge conjugation, hence the expectation value of 
the electric current (normal ordered w.r.t. the Minkowski vacuum) 
vanishes in the
groundstate (of the torus) and, therefore, the interaction Lagrangian 
${\cal L}$ (3.3) keeps the same form as on Minkowski space.
\vskip 1cm
{\bf Acknowledgements}: We profitted from discussions with Franz-Marc Boas, 
Izumi Ojima and Marek J. Radzikowski which are gratefully acknowledged.


\begin{thebibliography}{999}

\bibitem{BRS} Becchi, C., Rouet, A., and Stora, R., 
{\it Commun. Math. Phys.} {\bf 42}, 127 (1975)\\
Becchi, C., Rouet, A., and Stora, R.,
{\it Annals of Physics (N.Y.)} {\bf 98}, 287 (1976)

\bibitem{BlSe}Blanchard, P., and Seneor, R.,
{\it Ann. Inst. H. Poincar\'e A} {\bf 23}, 147 (1975)

\bibitem{BDF}Boas, F.M., D\"utsch, M., and K.Fredenhagen, K.,"A local 
(perturbative) construction 
of observables in gauge theories: nonabelian gauge theories", work in progress

\bibitem{BS}Bogoliubov, N.N., and Shirkov, D.V., {\it "Introduction to the 
Theory of Quantized Fields"}, New York (1959)

\bibitem{BW}Bordemann, M., and Waldmann, S.,
q-alg/9611004, to appear in {\it Commun. Math. Phys.}

\bibitem{BF}Brunetti, R., and Fredenhagen, K.,
"Interacting quantum fields in curved space: Renormalization of $\phi^4$", 
gr-qc/9701048, {\it Proceedings of the Conference 'Operator Algebras 
and Quantum Field Theory', held at Accademia Nazionale dei Lincei, 
Rome, July 1996}.\\
Brunetti, R., and Fredenhagen, K.,
"Microlocal analysis and interacting quantum field 
theories: Renormalization on physical backgrounds", in preparation.

\bibitem{BFK}Brunetti, R., Fredenhagen, K., and K\"ohler, M.,
{\it Commun. Math. Phys.} {\bf 180}, 312 (1996)

\bibitem{DF}D\"utsch, M., and Fredenhagen, K.,
"A local (perturbative) construction of observables 
in gauge theories: the example of QED", 
preprint: hep-th/9807078, DESY 98-090

\bibitem{DHS}D\"utsch, M., Hurth, T., and Scharf, G., 
{\it N. Cimento A} {\bf 108}, 737 (1995)

\bibitem{DHKS}D\"utsch, M., Hurth, T., Krahe, F., and Scharf, G.,
{\it N. Cimento A} {\bf 106}, 1029 (1993)

\bibitem{DKS}D\"utsch, M., Krahe, F., and Scharf, G.,
{\it N. Cimento A} {\bf 103}, 871 (1990)

\bibitem{EG}Epstein, H., and Glaser, V.,
{\it Ann. Inst. H. Poincar\'e A} {\bf 19}, 211 (1973)

\bibitem{K}Krahe, F.,
{\it Acta Phys. Polonica B} {\bf 27}, 2453 (1996)

\bibitem{KO}Kugo, T., and Ojima, I., 
{\it Suppl. Progr. Theor. Phys.} {\bf 66}, 1 (1979)

\bibitem{S}Scharf, G.,
{\it "Finite Quantum Electrodynamics. The causal approach"}, 
2nd. ed., Springer-Verlag (1995)

\bibitem{St1}Stora, R.,
"Lagrangian field theory", 
summer school of theoretical physics about {\it "particle physics"}, 
Les Houches, 1-79 (1971)

\bibitem{St2}Stora, R.,
"Differential algebras in Lagrangean field theory", 
ETH-Z\"urich Lectures, January-February 1993;\\
Popineau, G., and Stora, R.,
"A pedagogical remark on the main theorem of perturbative 
renormalization theory", unpublished preprint (1982)

\end{thebibliography}
\end{document}